\documentclass[12pt]{article}

\usepackage{graphicx, epsfig}
\usepackage{amssymb, bm}
\usepackage[centertags]{amsmath}
\usepackage[english]{babel}
\usepackage[authoryear]{natbib}

\newcommand{\Es}{\mathbb{E}}

\newcommand{\tr}{{\rm tr}}

\title{Small-sample corrections for score tests in Birnbaum--Saunders regressions}
\author{Artur J.~Lemonte, \quad Silvia L. P.~Ferrari\\
{\small {\em Departamento de Estat\'istica, Universidade de S\~ao Paulo, Rua do Mat\~ao, 1010,}}\\
{\small {\em S\~ao Paulo/SP, 05508-090, Brazil}}}
\date{}

\begin{document}
\maketitle

\begin{abstract}

In this paper we deal with the issue of performing accurate small-sample inference 
in the Birnbaum--Saunders regression model, which can be useful for modeling
lifetime or reliability data. We derive a Bartlett-type correction for the score test 
and numerically compare the corrected test with the usual score test, 
the likelihood ratio test and its Bartlett-corrected version.
Our simulation results suggest that the corrected test we propose
is more reliable than the other tests.\\


\end{abstract}

\section{Introduction}

The Birnbaum-Saunders distribution, also known as the fatigue life distribution, 
was introduced by \cite{BSa1969} and has received considerable attention in recent 
years. It was originally derived from a model for a physical fatigue process where dominant 
crack growth causes failure. It was later derived by \cite{Desmond85} using a
biological model which followed from relaxing some of the assumptions
originally made by \cite{BSa1969}.

The random variable $T$ is said to have a Birnbaum--Saunders distribution with
parameters $\alpha, \eta > 0$, say $\mathcal{B}$-$\mathcal{S}(\alpha, \eta)$,
if its cumulative distribution function (cdf) is given by
\[
F_{T}(t)=\Phi\biggl[\frac{1}{\alpha}\biggl(\sqrt{\frac{t}{\eta}}
        -\sqrt{\frac{\eta}{t}}\biggr)\biggr],\quad t > 0,
\]
where $\Phi(\cdot)$ is the standard normal distribution function
and $\alpha$ and $\eta$ are shape and scale parameters,
respectively. This distribution has a number of interesting properties:
(i) $\eta$ is its median; 
(ii) for any $k>0$, $kT \sim\mathcal{B}$-$\mathcal{S}(\alpha, k\eta)$;
(iii) its hazard function equals zero at $t = 0$, increases up to 
a maximum value and then decreases towards a given positive level 
\citep[see][]{KNBala};
(iv) $Y=\log(T)$ is sinh-normal distributed, say $Y\sim\mathcal{SN}(\alpha,\mu,\sigma)$, with shape, location and 
scale parameters given by $\alpha$, $\mu=\log(\eta)$ and
$\sigma=2$, respectively (see Section \ref{regBS}).
\cite{Rieck99} derived the moment generating function of the 
sinh-normal distribution and showed that it can be used to obtain 
both integer and fractional moments for the
$\mathcal{B}$-$\mathcal{S}(\alpha, \eta)$ distribution.

\cite{Rieck95} derived estimators for the parameters of the $\mathcal{B}$-$\mathcal{S}(\alpha, \eta)$ distribution 
in complete samples and type II symmetrically censored samples.
Some interesting results about improved statistical inference for the
$\mathcal{B}$-$\mathcal{S}(\alpha, \eta)$ distribution are available
in \cite{LCNV07}. Extensions of the Birnbaum--Saunders distribution 
are presented in \cite{SNL08}, \cite{Diaz-Leiva05}
and \cite{GPB09}. \cite{RiekNedelman91} proposed a log-linear regression 
model based on the Birnbaum--Saunders distribution. 
Diagnostic tools for the Birnbaum--Saunders regression model were developed by \cite{GLP04}, 
\cite{LBPG07} and \cite{XiWei07}, and Bayesian inference was developed by \cite{Tisionas01}.

Inference in Birnbaum--Saunders regressions relies on large-sample theory. 
For instance, a $\chi^2$ approximation to the distribution of the likelihood ratio
statistic can be used to perform approximate inference if the sample
size is large. \cite{Lemonte-et-al-2010} presented simulation results showing that the
(asymptotic) likelihood ratio test can be markedly oversized in small or moderate-sized 
samples. They derived Bartlett-corrected likelihood ratio tests that perform much better than the
original test. 

Rao's score test is an alternative to the likelihood ratio test. It is simpler 
to use since it only requires estimation under the null hypothesis. Also, it 
is often less size distorted than the likelihood ratio test; see the simulation
results in \cite{CBBF03}. Like the likelihood ratio test, the score test uses an
asymptotic approximation to the null distribution of the test statistic. This
approximation can be improved by applying a Bartlett-type correction to the score
statistic; see \cite{Cordeiro-Ferrari1991} for details. The correction needs to be
tailored for each application of interest. In this paper, our focus is the
Birnbaum--Saunders regression model. 

The paper is organized as follows. In Section~\ref{regBS} the Birnbaum--Saunders
regression model is presented. In Section~\ref{cumulants} we present some cumulants
of log-likelihood derivatives, which are needed in the subsequent section. In
Section~\ref{bartlett-type} we derive a Bartlett-type correction for the score
statistic. Numerical results are presented and discussed in Section~\ref{simulations}. 
We compare the performance of the test 
that uses the corrected score statistic with tests that use one of the following
statistics: score, likelihood ratio and its Bartlett-corrected version.
Our Monte Carlo simulations suggest that the test that uses the corrected score
statistic is much more accurate than the rival tests in small and moderate-sized
samples. Finally, Section~\ref{conclusions} closes the paper with some conclusions.

\section{The Birnbaum--Saunders regression model}\label{regBS}

Let $T\sim\mathcal{B}$-$\mathcal{S}(\alpha, \eta)$. The
density function of $Y=\log(T)$ is 
\[
f_{Y}(y)=\frac{2}{\alpha\sigma\sqrt{2\pi}}\cosh\biggl(\frac{y -
\mu}{\sigma}\biggr)\exp\biggl\{-\frac{2}{\alpha^2}
\mathrm{sinh}^2\biggl(\frac{y-\mu}{\sigma}\biggr)\biggr\}, \quad y\in\mathrm{I\!R}
\]
and, as before, we write $Y\sim\mathcal{SN}(\alpha,\mu,\sigma)$. 
\cite{RiekNedelman91} proposed the following regression model: 
\begin{equation}\label{eq1}
y_{i} = \bm{x}_{i}^{\top}\bm{\beta} + \varepsilon_{i},\quad i = 1, 2, \ldots, n,
\end{equation}
where $y_{i}$ is the logarithm of the $i$th observed lifetime,
$\bm{x}_{i}^{\top} = (x_{i1}, x_{i2}, \ldots, x_{ip})$ contains the $i$th
observation on the $p$ covariates ($p < n$),
$\bm{\beta} = (\beta_1, \beta_2, \ldots, \beta_p)^{\top}$ is a vector of
unknown regression parameters, and $\varepsilon_{i}\sim\mathcal{SN}(\alpha, 0, 2)$.
The distribution of the errors $\varepsilon_{i}$ has the following properties:
(i) it is symmetric around zero; (ii) it is unimodal for $\alpha\leq 2$ and bimodal for $\alpha > 2$; 
(iii) its variance is a function of $\alpha$ only, and has no closed-form expression, but \cite{Rieck89} 
obtained asymptotic approximations for both small and large values of $\alpha$;
(iv) $\alpha^{-1}\varepsilon_{i}$ converges in distribution to the
standard normal distribution when $\alpha\to 0$.

The log-likelihood function for the parameter vector  $\bm{\theta}=(\bm{\beta}^{\top},\alpha)^{\top}$,
apart from an unimportant constant, can be expressed as
\begin{equation}\label{eq2}
\ell(\bm{\theta})=\sum_{i=1}^{n}\log(\xi_{i1}) - \frac{1}{2}\sum_{i=1}^{n}\xi_{i2}^{2},
\end{equation}
where $\xi_{i1}=\xi_{i1}(\bm{\theta})= 2\alpha^{-1}\cosh([y_i-\mu_{i}]/2)$,
$\xi_{i2}=\xi_{i2}(\bm{\theta})=2\alpha^{-1}\sinh([y_i-\mu_{i}]/2)$
and $\mu_{i} = \bm{x}_{i}^{\top}\bm{\beta}$, for $i=1,2,\ldots,n$.
The $n\times p$ matrix $\bm{X}=(\bm{x}_{1}, \bm{x}_{2}, \ldots, \bm{x}_{n})^{\top}$
is assumed to be of full rank, i.e., rank($\bm{X})=p$.

\section{Cumulants of log-likelihood derivatives}\label{cumulants}

Derivatives of $\ell(\bm{\theta})$ with respect to the components of $\bm{\beta}$
and $\alpha$ are denoted by: $U_{r} = \partial\ell(\bm{\theta})/\partial\beta_{r}$,
$U_{\alpha}=\partial\ell(\bm{\theta})/\partial\alpha$,
$U_{rs}=\partial^{2}\ell(\bm{\theta})/\partial\beta_{r}\partial\beta_{s}$,
$U_{r\alpha}=\partial^{2}\ell(\bm{\theta})/\partial\beta_{r}\partial\alpha$,
$U_{rst}=\partial^{3}\ell(\bm{\theta})/\partial\beta_{r}\partial\beta_{s}\partial\beta_{t}$,
$U_{rs\alpha}=\partial^{3}\ell(\bm{\theta})/\partial\beta_{r}\partial\beta_{s}\partial\alpha$, etc.
Further, we use the following notation for joint cumulants of log-likelihood
derivatives: $\kappa_{rs} = \Es(U_{rs})$, $\kappa_{r,\alpha} = \Es(U_{r}U_{\alpha})$,
$\kappa_{rst} = \Es(U_{rst})$,
$\kappa_{rs,\alpha\alpha} = \Es(U_{rs}U_{\alpha\alpha}) - \kappa_{rs}\kappa_{\alpha\alpha}$,
$\kappa_{r,s,t\alpha} = \Es(U_{r}U_{s}U_{t\alpha}) - \kappa_{r,s}\kappa_{t\alpha}$,
etc. All $\kappa$'s are assumed to be of order $\mathcal{O}(n)$.
By differentiating~(\ref{eq2}) we have
\[
U_{r}=\frac{1}{2}\sum_{i=1}^{n}x_{ir}\biggl(\xi_{i1}\xi_{i2} - \frac{\xi_{i2}}{\xi_{i1}}\biggr)\quad
{\rm and}\quad
U_{\alpha}=-\frac{n}{\alpha} + \frac{1}{\alpha}\sum_{i=1}^{n}\xi_{i2}^{2}.
\]
The score function for $\bm{\beta}$ is
$\bm{U}_{\bm{\beta}}=(1/2)\bm{X}^{\top}\!\bm{s}$, where $\bm{s}=\bm{s}(\bm{\theta})$
is an $n$-vector whose $i$th element $\xi_{i1}\xi_{i2}-\xi_{i2}/\xi_{i1}$.

As will be seen below, some cumulants of log-likelihood derivatives depend on 
\[
a_{0,\alpha} = \biggl\{1 - \mathtt{erf}\biggl(\frac{\sqrt{2}}{\alpha}\biggr)\biggr\}
\exp\biggl(\frac{2}{\alpha^2}\biggr). 
\]
Here, ${\tt erf}(\cdot)$ represents the {\em error function}. Details on
the error function can be found in \cite{GR07}.
For small values of $\alpha$ \citep[p.~298]{AE70}
\begin{equation}\label{a0approx}
a_{0,\alpha} \approx \frac{\alpha}{\sqrt{2\pi}}\biggl\{1 - \frac{\alpha^2}{4} +
\frac{3\alpha^4}{16}\biggr\}. 
\end{equation}
For numerical evaluation we recommend the use of~(\ref{a0approx}) when $\alpha < 0.5$.

The complete list of the cumulants of log-likelihood derivatives that are 
needed in the derivation of Bartlett-corrected score statistic in the
Birnbaum--Saunders regression model is given below. Some cumulants were obtained by
\cite{Lemonte-et-al-2010}.
Using Bartlett identities, we arrive, after long algebra, at
$\kappa_{rst} = \kappa_{rs,t} = \kappa_{r,s,t} = 0$,
\[
\kappa_{rs} = -\frac{a_{1,\alpha}}{4}\sum_{i=1}^{n}x_{ir}x_{is},\quad
\kappa_{r,s,tu} = -\frac{s_{1,\alpha}}{2}\sum_{i=1}^{n}x_{ir}x_{is}x_{it}x_{iu}
\]
and
\[
\kappa_{r,s,t,u} = \frac{3s_{1,\alpha}}{2}\sum_{i=1}^{n}x_{ir}x_{is}x_{it}x_{iu},
\]
where $a_{1,\alpha} = 2 + 4/\alpha^{2} - a_{0,\alpha}\sqrt{2\pi}/\alpha$,
$s_{1,\alpha} = -2a_{2,\alpha}+(s_{0,\alpha} - a_{1,\alpha}^2)/8$,
$a_{2,\alpha} = -\{2 + 7/\alpha^2 -a_{0,\alpha}\sqrt{\pi}(1/(2\alpha)+ 6/\alpha^3)/\sqrt{2}\}/4$
and $s_{0,\alpha} = 12 + 2/\alpha^2 + 16/\alpha^4 +a_{0,\alpha}\sqrt{\pi}(1/\alpha + 12/\alpha^3)/\sqrt{2}$.
Also, $\kappa_{r\alpha\alpha} = \kappa_{r,\alpha\alpha} = \kappa_{r,\alpha,\alpha} = \kappa_{\alpha,\alpha r} = 0$,
\[
\kappa_{\alpha\alpha} = -\frac{2n}{\alpha^2},\quad
\kappa_{rs\alpha} = -\kappa_{r,s\alpha} = \frac{(2+\alpha^2)}{\alpha^3}\sum_{i=1}^{n}x_{ir}x_{is},\quad
\kappa_{\alpha\alpha,\alpha} = -\frac{6n}{\alpha^3},\quad
\kappa_{\alpha\alpha\alpha} = \frac{10n}{\alpha^3},
\]
\[
\kappa_{r,s,\alpha} = \biggl\{\frac{2(2+\alpha^2)}{\alpha^3} - a_{3,\alpha}\biggr\}\sum_{i=1}^{n}x_{ir}x_{is},\quad
\kappa_{rs,\alpha} = \biggl\{a_{3,\alpha} - \frac{(2+\alpha^2)}{\alpha^3}\biggr\}\sum_{i=1}^{n}x_{ir}x_{is},
\]
with $a_{3,\alpha} = 3/\alpha^3 -a_{0,\alpha}\sqrt{2\pi}(1/(4\alpha^2) + 1/\alpha^4)$.
Additionally,
\[
\kappa_{\alpha,\alpha,\alpha} = \frac{8n}{\alpha^3},\quad
\kappa_{r,s,\alpha\alpha} = -\frac{3}{\alpha^2}
\biggl\{\frac{2+\alpha^2}{\alpha^2} + \frac{s_{2,\alpha} -
a_{1,\alpha}}{4}\biggr\}\sum_{i=1}^{n}x_{ir}x_{is},
\]
\[
\kappa_{\alpha,\alpha,\alpha,\alpha} = \frac{48n}{\alpha^4},\quad
\kappa_{r,s,\alpha,\alpha} = \biggl\{\frac{2+11\alpha^2}{\alpha^4}  +
\frac{3(s_{2,\alpha} - a_{1,\alpha})}{4\alpha^2}- a_{4,\alpha}\biggr\}\sum_{i=1}^{n}x_{ir}x_{is},
\]
where $s_{2,\alpha} = 6 + 8a_{0,\alpha}\sqrt{\pi}/(\alpha^3\sqrt{2})$
and $a_{4,\alpha} = -10/\alpha^4 -4/\alpha^6 + a_{0,\alpha}\sqrt{\pi}
(\alpha^4 + 10\alpha^2 + 8)/(\alpha^7\sqrt{2})$.

\section{Improved Score Tests}\label{bartlett-type}
                            
The hypothesis of interest is
$\mathcal{H}_{0}\!\!: \bm{\beta}_{1} = \bm{\beta}_{1}^{(0)}$,
which will be tested against the alternative hypothesis
$\mathcal{H}_{1}\!\!:\bm{\beta}_{1}\neq\bm{\beta}_{1}^{(0)}$, 
where $\bm{\beta}$ is partitioned as $\bm{\beta} = (\bm{\beta}_{1}^{\top},
\bm{\beta}_{2}^{\top})^{\top}$, with
$\bm{\beta}_{1} = (\beta_{1}, \beta_{2},\dots,\beta_{q})^{\top}$ and
$\bm{\beta}_{2} = (\beta_{q+1}, \beta_{q+2},\dots,\beta_{p})^{\top}$ for $q < p$. Here,
$\bm{\beta}_{1}^{(0)}$ is a fixed column vector of dimension $q$. When $q = p$,
the null hypothesis is $\mathcal{H}_{0}\!\!: \bm{\beta} = \bm{\beta}^{(0)}$. 
The score statistic can be written as
\[
S_{\!R} = \widetilde{a}_{1,\alpha}^{-1}\widetilde{\bm{s}}^{\top}\bm{X}_{1}(\bm{R}^{\top}\bm{R})^{-1}
\bm{X}_{1}^{\top}\widetilde{\bm{s}},
\]
where $a_{1,\alpha}$ and $\bm{s}$ were defined above,
$\bm{R} = \bm{X}_{1} - \bm{X}_{2}(\bm{X}_{2}^{\top}\bm{X}_{2})^{-1}
\bm{X}_{2}^{\top}\bm{X}_{1}$, with the matrix $\bm{X}$ partitioned
as $\bm{X} = [\bm{X}_{1}\ \bm{X}_{2}]$ when $q < p$, following the partition of
$\bm{\beta}$. When $q = p$, $\bm{X} = \bm{X}_{1}$. Finally,
tilde indicates evaluation at the restricted maximum
likelihood estimate under the null hypothesis. 

Our aim is to obtain simple formulas for $A_{1}$, $A_{2}$ and $A_{3}$, the 
coefficients of order $n^{-1}$ that define the Edgeworth expansion of $S_{\!R}$
under the null hypothesis. These quantities also define the Bartlett-type correction for
the score statistic given by \cite{Cordeiro-Ferrari1991}. 
The Bartlett-type corrected score statistic is given by 
\begin{equation}\label{SR*}
S_{\!R}^{*} = S_{\!R}\biggl(1 - \sum_{k=1}^{3}\vartheta_{k}S_{\!R}^{k-1}\biggr),
\end{equation}
with
$\vartheta_{1} = (A_{1} - A_{2} + A_{3})/12q$, 
$\vartheta_{2} = (A_{2} - 2A_{3})/\{12q(q+2)\}$ and
$\vartheta_{3} = A_{3}/\{12q(q+2)(q+4)\}$. 
General expressions for the $A$'s are given by \cite{Harris85}. 
The coefficients $A_{1}$, $A_{2}$ and $A_{3}$ are functions of joint 
cumulants of log-likelihood derivatives up to the fourth order. Whenever they 
depend on unknown parameters, they can be evaluated at the 
restricted maximum likelihood estimates under the null hypothesis.
The null distribution of $S_R^*$ is chi-square with approximation error 
reduced from order $\mathcal{O}(n^{-1})$ to $\mathcal{O}(n^{-3/2})$.
The improved statistic $S^\ast_R$ can be nonmonotonic 
(in the unmodified statistic).  Alternative forms of the Bartlett-type
corrected score statistic that are monotonic transformations of $S_R$
can be found in \cite{Kakizawa1996} and \cite{CFC1998}.
For a detailed survey of Bartlett and Bartlett-type corrections in econometrics and
statistics, the reader is referred to \cite{CribariNetoCordeiro96}.

In the Birnbaum--Saunders regression model, $\alpha$ and $\bm{\beta}$ are globally
orthogonal, i.e.~$\kappa_{r,\alpha} = 0$, for $r=1,2,\ldots,p$. This fact yields some
simplification in the derivation of the $A$'s. In this case, we can write 
$A_{1} = A_{1,\bm{\beta}} + A_{1,\bm{\beta}\alpha}$, $A_{2} = A_{2,\bm{\beta}} +
A_{2,\bm{\beta}\alpha}$ and $A_{3} = A_{3,\bm{\beta}} + A_{3,\bm{\beta}\alpha}$,
where $A_{1,\bm{\beta}}$, $A_{2,\bm{\beta}}$ and $A_{3,\bm{\beta}}$ are the
corresponding $A$'s obtained as if $\alpha$ were known; the additional terms, $A_{1,\bm{\beta}\alpha}$,
$A_{2,\bm{\beta}\alpha}$ and $A_{3,\bm{\beta}\alpha}$, represent the contributions 
due to the fact that $\alpha$ is unknown. 
The additive structure of the $A$'s also holds for other 
models, such as the generalized linear models \citep{CF1995} and
exponential family nonlinear models \citep{FUC1997}.
It can be shown that
$A_{3,\bm{\beta}\alpha} = 0$.
Also, general expressions for $A_{1,\bm{\beta}\alpha}$ and $A_{2,\bm{\beta}\alpha}$ are given in the Appendix.

Let $\bm{Z} = \bm{X}(\bm{X}^{\top}\bm{X})^{-1}\bm{X}^{\top}$,
and for $q < p$, $\bm{Z}_{2} = \bm{X}_{2}(\bm{X}_{2}^{\top}\bm{X}_{2})^{-1}\bm{X}_{2}^{\top}$. 
Consider $\bm{Z}^{(2)} = \bm{Z}\odot\bm{Z}$, $\bm{Z}_{d}^{(2)} = \bm{Z}_{d}\odot\bm{Z}_{d}$,
etc., $\odot$ denoting the Hadamard (elementwise)
product of matrices and the subscript $d$ indicates that the off-diagonal elements 
of the matrix were set equal to zero. After long algebra, we obtain
\begin{equation}\label{Abeta}
A_{1,\bm{\beta}} = g_{1,\alpha}\tr\{(\bm{Z} - \bm{Z}_{2})_{d}\bm{Z}_{2d}\},\quad
A_{2,\bm{\beta}} = g_{2,\alpha}\tr\{(\bm{Z} - \bm{Z}_{2})_{d}^{(2)}\},\quad 
A_{3,\bm{\beta}} = 0,
\end{equation}
where $\tr(\cdot)$ represents the trace operator,
$g_{1,\alpha} = -96s_{1,\alpha}/a_{1,\alpha}^2$ and
$g_{2,\alpha} = 72s_{1,\alpha}/a_{1,\alpha}^2$.
When $q = p$, we have $\bm{Z}_{2} = \bm{0}$ and, hence, 
$A_{1,\bm{\beta}} = A_{3,\bm{\beta}} = 0$
and $A_{2,\bm{\beta}}$ can be written as
$A_{2,\bm{\beta}} = g_{2,\alpha}\tr\{\bm{Z}_{d}^{(2)}\}$.
Notice that the formulas given in~(\ref{Abeta}) are functions of
$(\bm{Z} - \bm{Z}_{2})_{d}$, $\bm{Z}_{2d}$ and $\alpha$; they do not
depend on the unknown parameter vector $\bm{\beta}$. Also,
notice that $A_{3,\bm{\beta}} = 0$.
The formulas in  (\ref{Abeta}) involve simple matrix algebra and
can be easily implemented in existing software such as {\tt R} \citep{CoreTeam}, {\tt Ox}
\citep{DcK}, {\tt MAPLE} \citep{Abell-et-al-2002}, among others.

We obtain
\begin{equation}\label{Abeta-alpha}
A_{1,\bm{\beta}\alpha} = \frac{12q}{n}\bigl\{(p-q)g_{4,\alpha} + g_{5,\alpha} + g_{6,\alpha}\bigr\},\quad
A_{2,\bm{\beta}\alpha} = \frac{q(q+2)g_{3,\alpha}}{n},
\end{equation}
where $g_{3,\alpha} = -24\alpha^2s_{3,\alpha}^2/a_{1,\alpha}^2$,
$g_{4,\alpha} = 4(2+\alpha^2)s_{3,\alpha}/(\alpha^2a_{1,\alpha}^2)$,
$g_{5,\alpha} = \alpha s_{3,\alpha}/a_{1,\alpha}$,
$g_{6,\alpha} = -\alpha^2 s_{4,\alpha}/a_{1,\alpha}$, 
$s_{3,\alpha} = 2(2+\alpha^2)/\alpha^3 - a_{3,\alpha}$,
$s_{4,\alpha} = -\{4(1-2\alpha^2)/\alpha^4 +  a_{4,\alpha}\}$.
It is interesting to note that $A_{1,\bm{\beta}\alpha}$ and $A_{2,\bm{\beta}\alpha}$
given in~(\ref{Abeta-alpha}) depend neither on the model matrix $\bm{X}$ 
(except through its rank) nor on the unknown parameter vector $\bm{\beta}$.
It is noteworthy that these formulas are very simple and of easy computational
implementation.

We now focus on testing the null hypothesis
$\mathcal{H}_{0}\!\!:\alpha=\alpha^{(0)}$ against the alternative
hypothesis $\mathcal{H}_{1}\!\!:\alpha\neq\alpha^{(0)}$,
where $\alpha^{(0)}$ is a known positive constant. The score statistic
for testing $\mathcal{H}_{0}$ can be written as
\[
S_{\!R} = \frac{n}{2}(\widetilde{\xi}_{2} - 1)^{2},
\]
where $\widetilde{\xi}_{2} = \sum_{i=1}^{n}\widetilde{\xi}_{i2}^{2}/n$.
After some algebra, we have
\[
A_{1} = \frac{24p}{n\alpha^4a_{1,\alpha}^2}\bigl\{(2+\alpha^2)^2(p+6)
- 4\alpha^3(2+\alpha^2)a_{3,\alpha} - \alpha^2(4+5\alpha^2)a_{1,\alpha}\bigr\},\\
\]
\[
A_{2} = \frac{12}{n}\biggl\{3 - \frac{4(2+\alpha^2)p}{\alpha^2a_{1,\alpha}}\biggr\}
\quad{\rm and}\quad
A_{3} = \frac{40}{n}.
\]
Again, the formulas for the $A$'s are very simple, depend on $\bm{X}$
only through its rank and do not depend on the unknown parameter $\bm{\beta}$. 
Clearly, the $A$'s should be evaluated at $\alpha^{(0)}$.

\section{Monte Carlo simulations}\label{simulations}

We shall now report Monte Carlo simulation results on the finite sample behavior of
four tests in the Birnbaum--Saunders regression model, namely: the likelihood ratio
test ($LR$), the Bartlett-corrected likelihood ratio test 
($LR_{b}$),\footnote{$LR_{b} = LR/\{1 + B(\bm{\theta})/q\}$, where 
$B(\bm{\theta})$ is the Bartlett factor; see 
\cite{Lemonte-et-al-2010} for details.} the original score test ($S_{\!R}$) and the
Bartlett-corrected score test ($S_{\!R}^{*}$). We consider the model 
\[
y_{i} = \beta_{1}x_{i1} + \beta_{2}x_{i2} + \cdots + \beta_{p}x_{ip} + \varepsilon_{i},
\]
where $x_{i1} = 1$ and $\varepsilon_{i}\sim\mathcal{SN}(\alpha, 0, 2)$, 
$i = 1, 2, \ldots, n$.
The covariate values were selected as random draws from the $\mathcal{U}(0,1)$
distribution. The number of Monte Carlo replications was 10,000, the nominal levels
of the tests were $\gamma$ = 10\% and 5\%, and all simulations were carried
out using the \texttt{Ox} matrix programming language \citep{DcK}.  

At the outset, the null hypothesis is 
$\mathcal{H}_{0}\!\!:\beta_{p-1} =\beta_{p} = 0$,
which is tested against a two-sided alternative, the sample size is $n=25$ and
$\alpha = 0.5$ and 1.0. Different values of $p$ were considered. The values of the
response were generated using $\beta_{1} = \beta_{2} =\cdots=\beta_{p-2} = 1$.
The null rejection rates of the four tests are presented in Table~\ref{tab1}.
\begin{table}[!htp]
\renewcommand{\arraystretch}{1.1}
\begin{center}
\footnotesize
\caption{Null rejection rates (\%); $\alpha$ = 0.5 and 1.0, with $n = 25$.}\label{tab1}
\begin{tabular}{c c c c c| c c c c}\hline                                            %
  & \multicolumn{8}{c}{$\alpha = 0.5$} \\\cline{2-9}                              
  & \multicolumn{4}{c|}{$\gamma = 10\%$} & \multicolumn{4}{c}{$\gamma = 5\%$} \\\cline{2-9}                              
  $p$   & $LR$  & $LR_{b}$  & $S_{\!R}$ & $S_{\!R}^{*}$ 
        & $LR$  & $LR_{b}$  & $S_{\!R}$ & $S_{\!R}^{*}$\\\hline                      
    3   & 12.96 &  9.99 &  9.89 &  9.55 &  6.96 & 5.16 &  4.66 & 4.91\\              %
    4   & 14.26 & 10.71 & 11.35 & 10.17 &  7.80 & 5.18 &  5.17 & 4.81\\              %
    5   & 15.63 & 10.75 & 12.57 &  9.93 &  8.79 & 5.39 &  5.89 & 4.87\\              %
    6   & 17.04 & 11.15 & 13.74 & 10.24 & 10.10 & 5.99 &  7.12 & 5.11\\              %
    7   & 19.02 & 11.94 & 15.38 & 10.54 & 11.40 & 6.26 &  7.83 & 5.38\\              %
    8   & 20.73 & 12.54 & 17.21 & 10.83 & 12.93 & 6.81 &  9.54 & 5.43\\              %
    9   & 22.76 & 13.30 & 18.77 & 10.84 & 14.41 & 6.88 & 10.59 & 5.22\\              %
   10   & 24.83 & 14.42 & 21.19 & 11.40 & 16.46 & 8.20 & 12.41 & 5.98\\ \hline       %
  & \multicolumn{8}{c}{$\alpha = 1.0$} \\\cline{2-9}                              
  & \multicolumn{4}{c|}{$\gamma = 10\%$} & \multicolumn{4}{c}{$\gamma = 5\%$} \\\cline{2-9}                              
  $p$   & $LR$  & $LR_{b}$  & $S_{\!R}$ & $S_{\!R}^{*}$ 
        & $LR$  & $LR_{b}$  & $S_{\!R}$ & $S_{\!R}^{*}$\\\hline                      %
    3   & 12.82 & 10.04 &  9.00 &  9.77 &  6.79 & 5.18 &  4.11 & 4.91\\              %
    4   & 14.00 & 10.68 &  9.88 & 10.02 &  7.67 & 5.20 &  4.27 & 4.77\\              %
    5   & 15.41 & 10.91 & 10.92 & 10.17 &  8.49 & 5.59 &  5.06 & 5.17\\              %
    6   & 16.76 & 11.29 & 12.27 & 10.53 &  9.76 & 5.96 &  5.82 & 5.23\\              %
    7   & 18.41 & 11.93 & 13.52 & 11.08 & 10.93 & 6.29 &  6.54 & 5.51\\              %
    8   & 20.65 & 12.54 & 15.60 & 10.92 & 12.68 & 6.79 &  8.01 & 5.65\\              %
    9   & 22.18 & 13.43 & 17.00 & 11.35 & 14.36 & 6.86 &  8.90 & 5.41\\              %
   10   & 24.65 & 14.61 & 19.43 & 11.85 & 16.09 & 8.22 & 10.93 & 6.23\\ \hline       %
\end{tabular}                     
\end{center}
\end{table}
\normalsize

Note that the likelihood ratio ($LR$) and the score ($S_{\!R}$) tests are
markedly liberal; the size distortion increases with the number of
regressors. For instance, when $p = 7$, $\alpha=0.5$ and $\gamma = 10\%$,
their rejection rates are 19.02\% and 15.38\%, respectively. It is noticeable
that the score test is less liberal than the
likelihood ratio test. Both corrected tests ($LR_{b}$ and $S_{\!R}^{*}$) are much
less size distorted than the uncorrected tests.
The best performing test is the Bartlett-corrected score test ($S_{\!R}^{*}$);
it displayed rejection rates closer to the nominal levels in all cases.
In the situation mentioned above, the null rejection rates of the 
corrected tests are 11.94\% ($LR_{b}$) and 10.54\% ($S_{\!R}^{*}$).
Similar results hold for $\alpha=1.0$.
More importantly, simulations carried out for a wide range of values of
$\alpha$ reveal an analogous pattern (results not shown here for the sake
of space).\footnote{We used the approximation in (\ref{a0approx}) for $a_{0,\alpha}$ when $\alpha<0.5$.}
For instance, when $n = 25$, $p=4$ and $\gamma=10\%$, the rejection rates
of  $\mathcal{H}_{0}\!\!:\beta_{3} =\beta_{4} = 0$ are 14.33\% ($LR$),
11.42\% ($LR_{b}$), 11.76\% ($S_{\!R}$) and 10.08\% ($S_{\!R}^{*}$)
for $\alpha = 0.1$, and  12.44\% ($LR$), 11.23\% ($LR_{b}$), 5.12\%
($S_{\!R}$) and 9.52\% ($S_{\!R}^{*}$) for $\alpha = 10$. We noticed
that the uncorrected score test becomes conservative when $\alpha$ is large.

Table~\ref{tab2} reports results for sample sizes ranging from 15 to 100,
$\alpha = 0.5$ and $p=7$. The null hypothesis under test is 
$\mathcal{H}_{0}\!\!:\beta_{6} =\beta_{7} = 0$. 
The uncorrected tests are remarkably oversized when the sample size is 
small. As expected, the null rejection rates of all tests approach the 
corresponding nominal levels as the sample size grows. 
Clearly, the corrected tests are less size distorted than the 
unmodified tests, and the test that uses
$S_{\!R}^{*}$ has superior behavior than the tests that use $LR$, $LR_{b}$ and
$S_{\!R}$ in all cases.
For example, when $n = 20$ and $\gamma = 5\%$, the null rejection rates are 
13.87\% ($LR$), 7.06\% ($LR_{b}$), 9.15\% ($S_{\!R}$) and 5.62\% ($S_{\!R}^{*}$). 
\begin{table}[!htp]
\renewcommand{\arraystretch}{1.1}                      
\begin{center}
\footnotesize
\caption{Null rejection rates (\%); $\alpha = 0.5$,
        $p = 7$ and different sample sizes.}\label{tab2}
\begin{tabular}{c c c c c| c c c c}\hline                                            %
  & \multicolumn{4}{c|}{$\gamma = 10\%$} & \multicolumn{4}{c}{$\gamma = 5\%$} \\\cline{2-9}                              
  $n$   & $LR$  & $LR_{b}$  & $S_{\!R}$ & $S_{\!R}^{*}$ 
        & $LR$  & $LR_{b}$  & $S_{\!R}$ & $S_{\!R}^{*}$\\\hline                      %
     15  & 27.93 & 16.03 & 21.27 & 12.96 & 19.22 & 9.44 & 12.01 & 7.32 \\            %
     20  & 21.93 & 13.03 & 17.29 & 11.10 & 13.87 & 7.06 &  9.15 & 5.62 \\            %
     30  & 17.28 & 11.55 & 14.53 & 10.56 & 10.13 & 5.69 &  7.51 & 5.16 \\            %
     40  & 15.26 & 10.87 & 13.19 & 10.45 &  8.59 & 5.58 &  6.78 & 5.18 \\            %
     50  & 14.10 & 11.05 & 12.74 & 10.75 &  8.18 & 5.59 &  6.64 & 5.34 \\            %
    100  & 11.55 &  9.97 & 10.81 &  9.87 &  6.03 & 5.01 &  5.48 & 5.00 \\ \hline
\end{tabular}
\end{center}
\end{table}
\normalsize

We shall now present simulations results on the powers of the tests.
We set $p = 3$, $\alpha = 0.5$ and $n$ = 30 and 50. Since the unmodified tests 
are considerably oversized, we have only considered the two corrected
tests. The rejection rates were obtained under the alternative hypothesis
$\beta_{2} = \beta_{3} = \delta$, with different values of $\delta$ ($\delta > 0$). 
Note that the two tests display similar powers. For instance, when $n = 50$, $\gamma = 10\%$
and $\delta = 0.5$, the nonnull rejection rates are 86.90\% ($LR_{b}$) and 86.70\%
($S_{\!R}^*$). Not surprisingly, the powers of the tests increase with $n$ and also
with $\delta$; see Table~\ref{tab3}. 
\begin{table}[!htp]
\renewcommand{\arraystretch}{1.1}
\begin{center}
\footnotesize
\caption{Nonnull rejection rates (\%); $\alpha = 0.5$,
        $p = 3$ and $n$ = 30 and 50.}\label{tab3}
\begin{tabular}{c c c c | c c }\hline                                                %
&  & \multicolumn{2}{c|}{$LR_{b}$} & \multicolumn{2}{c}{$S_{\!R}^{*}$} \\\cline{3-6} %
  $n$ & $\delta$ & 10\%  & 5\% & 10\%  & 5\% \\ \hline
   30 & 0.1 & 12.69 &  6.81 & 12.47 &  6.62 \\            
      & 0.3 & 31.68 & 20.90 & 31.29 & 20.30 \\       
      & 0.5 & 64.43 & 50.32 & 63.64 & 49.26 \\            
      & 0.7 & 88.54 & 80.50 & 88.26 & 79.68 \\ \hline       
   50 & 0.1 & 14.80 &  8.03 & 14.76 &  7.97  \\              
      & 0.3 & 48.87 & 35.28 & 48.65 & 34.88  \\              
      & 0.5 & 86.90 & 78.48 & 86.70 & 78.17  \\  
      & 0.7 & 99.03 & 97.54 & 99.05 & 97.50  \\ \hline                                
\end{tabular}                                    
\end{center}
\end{table}
\normalsize

We performed Monte Carlo simulations considering hypothesis testing on $\alpha$. 
To save space, the results are not shown. We noticed that the corrected tests 
display superior behaviour than the uncorrected tests. For example, 
when $n=30$, $p=4$, $\gamma = 10\%$ and $\mathcal{H}_{0}\!\!:\alpha =1.0$, we
obtained the following null rejection rates: 19.86\%
($LR$), 10.73\% ($LR_{b}$), 13.77\% ($S_{\!R}$) and 9.89\% ($S_{\!R}^{*}$).
Again, the best performing test is the Bartlett-corrected score test.

As noted by the referee, our results can also be used to invert the 
Edgeworth expansion of the score statistic null distribution function; 
see \cite{Harris85} and \cite{Cordeiro-Ferrari1991} for details. The 
test is then performed by comparing the unmodified statistic with a 
corrected critical value. To be specific, let $\gamma$ be the desired 
level of the test and ${q}_{1-\gamma}$ be the $1-\gamma$ quantile of 
the $\chi^2$ limiting distribution of the test statistic. The corrected 
critical value is 
\begin{equation}\label{q*}
q_{1-\gamma}^{*} = q_{1-\gamma}\biggl(1 + \sum_{k=1}^{3}\vartheta_{k}\;q_{1-\gamma}^{k-1}\biggr),
\end{equation}
where $\vartheta_{i}$, for $i=1,2,3$, are given just below (\ref{SR*}).
We shall now present some simulation evidence on the finite sample behavior of
the score test that uses the corrected critical value ($S_H$). 
For the sake of comparison, the new set of 
simulation results includes rejection rates of the other 
four tests ($LR$, $LR_{b}$, $S_R$ and $S_{R}^{*}$) and the score test 
that uses parametric bootstrap critical values ($S_{boot}$). The bootstrap correct 
test is performed as follows. First, one generates $B$ bootstrap samples (we
set $B=600$) from the assumed model with the parameters replaced by 
restricted estimates computed using the original sample, under 
$\mathcal{H}_{0}$, i.e. imposing the restrictions stated in the null 
hypothesis. Second, for each pseudo-sample, 
one computes the score statistic; $S_R^{b}$ denotes the score statistic 
for the $b$-th sample, $b = 1, 2, \ldots, B$. Third,
the $1- \gamma$ percentile of $S_R$ is estimated by 
$\widehat{q}_{1-\gamma}$, such that 
$\#\{S_R^{b} \leq  \widehat{q}_{1-\gamma}\}/B = 1 - \gamma$.
Finally, one rejects the null hypothesis if $S_R > \widehat{q}_{1-\gamma}$. 

Table~\ref{tab_new1} reports rejection rates of the different tests of the null
hypothesis $\mathcal{H}_{0}\!\!:\beta_{8} =\beta_{9} = 0$ for
$\alpha$ = 1.5, $n = 30$ and $p=9$. The score test that uses the corrected 
critical value (\ref{q*}) and the Bartlett-corrected score test have similar 
size properties and are less size distorted than both likelihood ratio tests 
(corrected and uncorrected) and the original score test. The test that 
uses the bootstrap corrected critical value performs very well although being
slightly conservative. Its main disadvantage over the other tests is the need 
of a computer intensive procedure.

\begin{table}[!htp]
\renewcommand{\arraystretch}{1.1}
\begin{center}
\footnotesize
\caption{Null rejection rates (\%); $\alpha$ = 1.5, $n = 30$, $p=9$.}\label{tab_new1}
\begin{tabular}{c c c}\hline
    & $\gamma = 10\%$ & $\gamma = 5\%$ \\\hline                 
    $LR     $     & 19.36   & 11.71\\
    $LR_{b} $     & 12.17   & 6.40 \\
    $S_{\!R}$     & 14.79   & 7.62 \\
    $S_{\!R}^{*}$ & 11.25   & 5.66 \\
    $S_{H}$       & 11.42   & 5.64 \\     
    $S_{boot}$    &  9.53   & 4.71 \\\hline                                                              
\end{tabular}                     
\end{center}
\end{table}
\normalsize

Finally, we end this section by reporting the first four moments
of $LR$, $LR_{b}$, $S_R$ and $S_{R}^{*}$ 
and the corresponding moments of the limiting $\chi^2$ distribution 
in the setting of Table~\ref{tab_new1}; see Table~\ref{tab_new2}.
Clearly, the best agreement between the true moments (obtained by simulation) 
and the moments of the limiting distribution is achieved by the Bartlett-corrected 
score statistic, $S_{R}^{*}$.  

\begin{table}[!htp]
\renewcommand{\arraystretch}{1.1}
\begin{center}
\footnotesize
\caption{Moments; $\alpha$ = 1.5, $n = 30$, $p=9$.}\label{tab_new2}
\begin{tabular}{l c c c c c}\hline
  & $LR$ & $LR_{b}$ & $S_{\!R}$ & $S_{\!R}^{*}$ & $\chi_{2}^{2}$ \\\hline
Mean     &  2.81 &  2.20  &  2.41 & 2.10  &  2.00 \\
Variance &  8.08 &  4.94  &  4.89 & 4.32  &  4.00 \\
Skewness &  2.17 &  2.17  &  1.59 & 1.98  &  2.00 \\     
Kurtosis & 10.93 & 10.88  &  6.46 & 9.19  &  9.00 \\\hline                                                              
\end{tabular}                     
\end{center}
\end{table}
\normalsize

\section{Conclusions}\label{conclusions}

We addressed the issue of performing testing inference in Birnbaum--Saunders
regressions when the sample size is small or moderate. We derived a Bartlett-type correction 
for the score test. We numerically compared  the behaviour of four tests, 
namely: the likelihood ratio test ($LR$), its Bartlett-corrected version ($LR_{b}$),
the score test ($S_{\!R}$) and its Bartlett-type
corrected counterpart ($S_{\!R}^{*}$). 
Also, we perfomed a small simulation study that includes
two score tests that use critical values that are approximated either
analytically or by a parametric bootstrap method.
The bootstrap-corrected test performs well, but requires a 
computer intensive procedure.
Overall, our numerical results favor the test obtained from applying a Bartlett-type 
correction to the score test statistic.
Therefore, we recommend the corrected score
test proposed in the present paper for practical applications.

\section*{Acknowledgments}

We thank an anonymous referee for helpful suggestions which improved the paper. 
We gratefully acknowledge grants from FAPESP and CNPq (Brazil).

{\small
\appendix
\section*{Appendix}

Since $\kappa_{r,\alpha} = 0$, for $r=1,2,\ldots, p$, the total
Fisher information matrix for $\bm{\theta}$, $\bm{K}(\bm{\theta})$, 
and its inverse, $\bm{K}(\bm{\theta})^{-1}$, are block diagonal. Using
the partition of $\bm{X}$ induced by the null hypothesis
$\mathcal{H}_{0}\!\!: \bm{\beta}_{1} = \bm{\beta}_{1}^{(0)}$, we can write
\[
\bm{K}(\bm{\theta}) = \begin{bmatrix}
                        \bm{K}_{11} & \bm{K}_{12} & \bm{0}\\
                        \bm{K}_{21} & \bm{K}_{22} & \bm{0}\\            
                        \bm{0} & \bm{0}& \kappa_{\alpha, \alpha}
                        \end{bmatrix}
                                     \quad{\rm and}\quad
\bm{K}(\bm{\theta})^{-1} = \begin{bmatrix}                   
                             \bm{K}^{11} & \bm{K}^{12} & \bm{0}\\
                             \bm{K}^{21} & \bm{K}^{22} & \bm{0}\\            
                             \bm{0} & \bm{0}& \kappa_{\alpha, \alpha}^{-1}
                             \end{bmatrix}.
\]
Let
\[
\bm{A}_{\bm{\beta}} = \begin{bmatrix}                          
                        \bm{0} & \bm{0}\\               
                        \bm{0} & \bm{K}_{22}^{-1}
                        \end{bmatrix},\quad
\bm{K}_{\bm{\beta}}^{-1} = \begin{bmatrix}                       
                               \bm{K}^{11} & \bm{K}^{12}\\               
                               \bm{K}^{21} & \bm{K}^{22}
                               \end{bmatrix},\quad
\bm{A} = \begin{bmatrix}                      
          \bm{A}_{\bm{\beta}} & \bm{0}\\               
          \bm{0} & \kappa_{\alpha,\alpha}^{-1}
          \end{bmatrix}
                         \quad {\rm and}\quad
\bm{M} = \begin{bmatrix}                        
          \bm{M}_{\bm{\beta}} & \bm{0}\\               
          \bm{0} & 0
          \end{bmatrix},
\]
where $\bm{M}_{\bm{\beta}} = \bm{K}_{\bm{\beta}}^{-1} - \bm{A}_{\bm{\beta}}$.
Let $m_{r\alpha}$ and $a_{r\alpha}$ be the $(r,p+1)$-th
elements of $\bm{M}$ and $\bm{A}$, respectively, and let $m_{\alpha\alpha}$ and
$a_{\alpha\alpha}$ be the $(p+1, p+1)$-th elements of $\bm{M}$ and $\bm{A}$, respectively.
We have $m_{r\alpha}= m_{\alpha r} = m_{\alpha\alpha} = 0$,
$a_{r\alpha} = a_{\alpha r} = 0$ $(r = 1, 2,\ldots,p)$ and $a_{\alpha\alpha} = \kappa_{\alpha,\alpha}^{-1}$.
Therefore, the general formulas for $A_{1,\bm{\beta}\alpha}$ and $A_{2,\bm{\beta}\alpha}$ are
\begin{align*}
A_{1,\bm{\beta}\alpha} &= 3\sideset{}{^{\prime}}\sum(\kappa_{\alpha\alpha k} + 2\kappa_{\alpha,\alpha k})(\kappa_{rst} + 2\kappa_{rs,t})a_{\alpha\alpha}a_{st}m_{kr}\\
&+3\sideset{}{^{\prime}}\sum(\kappa_{ijk} + 2\kappa_{i,jk})(\kappa_{r\alpha\alpha} + 2\kappa_{r\alpha,\alpha})a_{ij}a_{\alpha\alpha}m_{kr}\\
& +3\sideset{}{^{\prime}}\sum(\kappa_{\alpha\alpha k} + 2\kappa_{\alpha,\alpha k})(\kappa_{r\alpha\alpha} + 2\kappa_{r\alpha,\alpha})a_{\alpha\alpha}^{2}m_{kr}\\
&-6\sideset{}{^{\prime}}\sum(\kappa_{\alpha\alpha k} + 2\kappa_{\alpha,\alpha k})\kappa_{r,s,t}a_{\alpha\alpha}a_{kr}m_{st}\\
& -6\sideset{}{^{\prime}}\sum(\kappa_{ij\alpha} + 2\kappa_{i,j\alpha})\kappa_{\alpha,s,t}a_{ij}a_{\alpha\alpha}m_{st}
-6\sideset{}{^{\prime}}\sum(\kappa_{\alpha\alpha\alpha} + 2\kappa_{\alpha,\alpha\alpha})\kappa_{\alpha,s,t}a_{\alpha\alpha}^{2}m_{st}\\
& +6\sideset{}{^{\prime}}\sum(\kappa_{i,\alpha k} - \kappa_{i,\alpha,k})(\kappa_{r\alpha t} + 2\kappa_{r\alpha,t})a_{\alpha\alpha}a_{kt}m_{ir}\\
&+6\sideset{}{^{\prime}}\sum(\kappa_{i,j\alpha} - \kappa_{i,j,\alpha})(\kappa_{rs\alpha} + 2\kappa_{rs,\alpha})a_{js}a_{\alpha\alpha}m_{ir}\\
& +6\sideset{}{^{\prime}}\sum(\kappa_{i,\alpha\alpha} - \kappa_{i,\alpha,\alpha})(\kappa_{r\alpha\alpha} + 2\kappa_{r\alpha,\alpha})a_{\alpha\alpha}^{2}m_{ir}
-6\sideset{}{^{\prime}}\sum(\kappa_{i,j,\alpha,\alpha} + \kappa_{i,j,\alpha\alpha})a_{\alpha\alpha}m_{ij}
\end{align*}
and
\begin{align*}
A_{2,\bm{\beta}\alpha} &= -3\sideset{}{^{\prime}}\sum\kappa_{i,j,\alpha}\kappa_{\alpha,s,t}a_{\alpha\alpha}m_{ij}m_{st}
         +6\sideset{}{^{\prime}}\sum(\kappa_{\alpha\alpha k} + 2\kappa_{\alpha,\alpha k})\kappa_{r,s,t}a_{\alpha\alpha}m_{kr}m_{st}\\
         &-6\sideset{}{^{\prime}}\sum\kappa_{i,j,\alpha}\kappa_{r,s,\alpha}a_{\alpha\alpha}m_{ir}m_{js},
\end{align*}
where $\sum^{\prime}$ denotes summation over all the components of $\bm{\beta}$,
and $a_{ij}$ e $m_{ij}$ $(i,j = 1, 2,\ldots,p)$ are the $(i,j)$-th elements of $\bm{A}$ 
and $\bm{M}$, respectively.

Plugging the cumulants given in Section~\ref{bartlett-type} in the formulas for
$A_{1,\bm{\beta}\alpha}$ and $A_{2,\bm{\beta}\alpha}$, we obtain~(\ref{Abeta-alpha}). 
To save space, we will only show how to obtain
 $-6\sum^{\prime}(\kappa_{ij\alpha} + 2\kappa_{i,j\alpha})\kappa_{\alpha,s,t}a_{ij}
a_{\alpha\alpha}m_{st}$. The other terms can be obtained in a similar fashion. 
Notice that
\[
-6\sideset{}{^{\prime}}\sum(\kappa_{ij\alpha} + 2\kappa_{i,j\alpha})\kappa_{\alpha,s,t}a_{ij}a_{\alpha\alpha}m_{st}=
\frac{6(2+\alpha^2)s_{3,\alpha}a_{\alpha\alpha}}{\alpha^3}
\sideset{}{^{\prime}}\sum\sum_{l,m=1}^{n}x_{li}x_{lj}x_{ms}x_{mt}a_{ij}m_{st}.
\]
Inverting the order of summation and rearranging the terms, we have
\[
-6\sideset{}{^{\prime}}\sum(\kappa_{ij\alpha} + 2\kappa_{i,j\alpha})\kappa_{\alpha,s,t}a_{ij}a_{\alpha\alpha}m_{st}=
\frac{3(2+\alpha^2)s_{3,\alpha}}{n\alpha}
\sum_{l,m=1}^{n}\sideset{}{^{\prime}}\sum a_{ij}x_{li}x_{lj}\sideset{}{^{\prime}}\sum m_{st}x_{ms}x_{mt}.
\]
The terms $\sum^{\prime}a_{ij}x_{li}x_{lj}$ and $\sum^{\prime}m_{st}x_{ms}x_{mt}$
represent the elements $(l,l)$ and $(m,m)$ of $4\bm{Z}_{2}/a_{1,\alpha}$ and
$4(\bm{Z} - \bm{Z}_{2})/a_{1,\alpha}$, respectively. The matrices 
$\bm{Z} = \{z_{lm}\}$ and $\bm{Z}_{2} = \{z_{2lm}\}$
were defined in Section~\ref{bartlett-type}. Hence
\[
-6\sideset{}{^{\prime}}\sum(\kappa_{ij\alpha} + 2\kappa_{i,j\alpha})\kappa_{\alpha,s,t}a_{ij}a_{\alpha\alpha}m_{st}=
\frac{48(2+\alpha^2)s_{3,\alpha}}{n\alpha a_{1,\alpha}^2}
\sum_{l,m=1}^{n}z_{2ll}(z_{mm} - z_{2mm}).
\]
From $\sum_{l,m=1}^{n}z_{2ll}(z_{mm} - z_{2mm}) =
\tr(\bm{Z}_{2})\tr(\bm{Z} - \bm{Z}_{2}) = (p-q)q$,
we obtain
\[
-6\sideset{}{^{\prime}}\sum(\kappa_{ij\alpha} + 2\kappa_{i,j\alpha})\kappa_{\alpha,s,t}a_{ij}a_{\alpha\alpha}m_{st}=
\frac{48s_{3,\alpha}(2+\alpha^2)(p-q)q}{n\alpha a_{1,\alpha}^2}.
\]
}

\end{document}